\documentclass[twocolumn,aps]{revtex4}

\usepackage{graphicx} % Include figure files

\begin{document}

\title{Long-range coupling and scalable architecture for superconducting flux qubits}

\author{Austin G. Fowler$^{1}$, William F. Thompson$^{1}$, Zhizhong Yan$^{1}$, Ashley M. Stephens$^{2}$, B. L. T. Plourde$^{3}$ and Frank K. Wilhelm$^{1}$}
\affiliation{$^{1}$Institute for Quantum Computing,\\
University of Waterloo, Waterloo, ON, CANADA\\
$^{2}$Centre for Quantum Computer Technology,\\
University of Melbourne, Victoria, AUSTRALIA\\
$^{3}$Department of Physics, Syracuse University, Syracuse, NY,
USA.}

\date{\today}

\begin{abstract}
Constructing a fault-tolerant quantum computer is a daunting task.
Given any design, it is possible to determine the maximum error rate
of each type of component that can be tolerated while still
permitting arbitrarily large-scale quantum computation.  It is an
underappreciated fact that including an appropriately designed
mechanism enabling long-range qubit coupling or transport
substantially increases the maximum tolerable error rates of all
components.  With this thought in mind, we take the superconducting
flux qubit coupling mechanism described in \cite{Plou04} and extend
it to allow approximately 500 MHz coupling of square flux qubits, 50
$\mu$m a side, at a distance of up to several mm. This mechanism is
then used as the basis of two scalable architectures for flux qubits
taking into account crosstalk and fault-tolerant considerations such
as permitting a universal set of logical gates, parallelism,
measurement and initialization, and data mobility.
\end{abstract}

\maketitle

\section{Introduction}

The field of quantum computation is largely concerned with the
manipulation of two state quantum systems called qubits.  Unlike the
bits in today's computers which can be either 0 or 1, qubits can be
placed in arbitrary superpositions $\alpha|0\rangle +
\beta|1\rangle$, and entangled with each other. For a complete
review of the basic properties of qubits and quantum information,
see \cite{Niel00}.  The attraction of quantum computation lies in
the existence of quantum algorithms that are in some cases
exponentially faster than their best known classical equivalents.
Most famous are Shor's factoring algorithm \cite{Shor94b} and
Grover's search algorithm \cite{Grov96}.  There has also been
extensive work on using a quantum computer to simulate quantum
physics \cite{Feyn82,Lloy96,Bogh98,Sorn99,Byrn06}, an ongoing
exploration of adiabatic algorithms \cite{Farh01,Chil00,Banu06},
plus the discovery of quantum algorithms for differential equations
\cite{Szko04b}, finding eigenvalues \cite{Abra99b,Jaks03}, numerical
integration \cite{Abra99} and various problems in group theory
\cite{Kita96,Mosc99,Fenn05} and knot theory \cite{Subr02,Wocj06}.

Quantum systems suffer from decoherence, meaning their state rapidly
becomes unknowable through unwanted interaction with the
environment.  Flux qubit decoherence times of up to a few
microseconds have been demonstrated \cite{Bert05} versus
single-qubit gate times of order 10 ns and likely initial two-qubit
gate times of order a few tens of nanoseconds
\cite{Plou04,Hime06,vdPl07}. To perform long quantum computations,
quantum error correction will be required
\cite{Cald95,Stea96,DiVi06}. It has been shown that provided the
totality of decoherence and control errors is below some nonzero
threshold, and given an arbitrarily long time and an arbitrary large
number of qubits, an arbitrarily long and large quantum computation
can be performed \cite{Knil96b}. Despite being well known in certain
circles, the broader quantum computing community has not yet
sufficiently come to terms with the fact that long-range
interactions permit much higher levels of decoherence and control
error to be tolerated. With unlimited range interactions and
extremely large numbers of qubits, the threshold error rate has been
shown to be of order $10^{-2}$ \cite{Knil04c}.  With fewer qubits
but still unlimited range interactions, the threshold is reduced to
between $10^{-3}$ and $10^{-4}$ \cite{Stea03b}. A 2D lattice of
qubits interacting with their nearest neighbors only has been
devised with approximate threshold $10^{-5}$ \cite{Svor06}. The full
analysis of an infinite double line of qubits with nearest neighbor
interactions has been performed yielding a lower bound to the
threshold of $1.96 \times 10^{-6}$ \cite{Step07a}.  Work on an
infinite single line of qubits with nearest neighbor interactions is
in progress, and the threshold is expected to be of order $10^{-8}$
\cite{Step07b}.

Despite the extremely low expected threshold of linear nearest
neighbor (LNN) architectures, a great deal of theoretical work has
been devoted to the design of such architectures using a variety of
physical systems
\cite{Wu00,Vrij00,Gold03,Nova04,Holl03,Tian03,Frie03,Vand02,Soli03,Vyur00,Kame03}.
This is reasonable in the context of providing an experimental
starting point, but we believe the time has come to expect at least
a theoretical proposal for how long-range interactions or
long-distance qubit transport might be performed. Without this, it
is extremely difficult to argue the long-term viability of a given
system.  Furthermore, any proposed method of interaction or
transport must be able to be performed in parallel on a number of
pairs of qubits that grows linearly with the size of the computer to
permit the simultaneous application of error correction to a
constant fraction of the logical qubits in the computer.  By
contrast, a quantum computer based around a single, global, serial
interaction or transport mechanism, such as a single resonator
shared by all qubits in the computer, cannot simultaneously apply
quantum error correction to multiple logical qubits.  Such a quantum
computer could at best apply quantum error correction to each
logical qubit in turn.  As the number of logical qubits increases
and the amount of time between applications of error correction to a
given logical qubit increases, more errors accumulate and the
probability of successful error correction decreases.  Beyond a
certain amount of time between error correction applications, it is
overwhelmingly likely that every physical qubit comprising a given
logical qubit will have suffered an error, meaning no amount of
quantum error correction will successfully recover the original
logical state. Consequently, any quantum computer based around a
single, global, serial interaction or transport mechanism is not
scalable in the sense that it could never perform an arbitrarily
large quantum computation.

The purpose of this paper is to present a long-range coupling
mechanism for superconducting flux qubits that can be used to couple
many pairs of qubits together in parallel in a manner suited to the
construction of an arbitrarily large fault-tolerant quantum
computer.  In Section \ref{coupfluxqub} we review the coupling
mechanism of \cite{Plou04,Hime06} and extend it to allow long-range
coupling. In Section \ref{archit} we firstly describe a simple, yet
scalable, flux qubit architecture based on this interaction, but not
taking full advantage of it, then a more complicated architecture
with a better threshold error rate though much more difficult to
build. Finally, Section \ref{conc} concludes with a summary of
results and a description of further work.

\section{Coupling flux qubits}
\label{coupfluxqub}

\begin{figure}
\begin{center}
\resizebox{80mm}{!}{\includegraphics{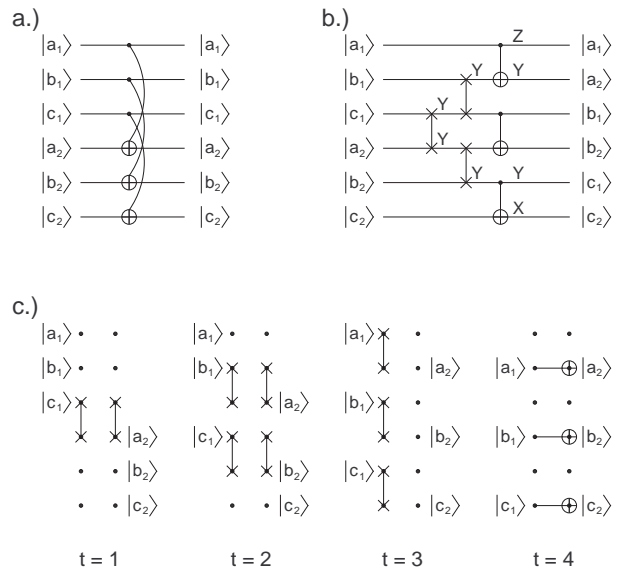}}
\end{center}
\caption{a.) nonlocal transversal interaction, b.) naive linear
nearest neighbor transversal interaction showing the propagation of
errors resulting from the failure of a single swap gate leading to
two errors in both logical qubits, c.) the first 4 time steps of a
bilinear fault-tolerant transversal interaction between two sets of
three 3 physical qubits. The remaining 3 steps are the time reverse
of the first 3 steps.} \label{comparison}
\end{figure}

Before discussing our coupling mechanism, a few words elaborating
exactly why long-range interactions are advantageous are in order.
Essentially, the problem lies in the need to perform transversal
multiple logical qubit gates, as shown in Fig.~\ref{comparison}a. If
long-range interactions are available, two logical qubits each
comprised of $n$ physical qubits can be transversely interacted in a
single time step using $n$ gates. If we now try to do the same thing
on a linear nearest neighbor architecture using swap gates prior to
the necessary gates to perform the transversal interaction, we
immediately run into a serious problem.  A single swap gate failure
can lead to two errors in a single logical qubit as shown in
Fig.~\ref{comparison}b. Under normal circumstances, two or more
errors in a single logical qubit are not correctable.  A solution to
this dilemma that avoids the need to resort to a multiple error
correcting code is shown in Fig.~\ref{comparison}c where a second
line of placeholder qubits has been added. Now swap gates never
simultaneously touch two qubits that are both part of logical
qubits.  Note that the first three steps of Fig.~\ref{comparison}c
need to be repeated, time reversed, to return the qubits to their
original configuration.  We can now see that to interact two
$n$-qubit logical qubits transversely and fault-tolerantly using
only nearest neighbor interactions, using the scheme described, we
need $4n$ qubits, $2n^2+n$  gates and $2n+1$ time steps with $2n^2$
locations where data qubits are left idle. For all of this
additional machinery to result in a circuit with the same
reliability as the nonlocal case, every individual component must be
significantly more reliable. This is the origin of the lower
thresholds quoted in the introduction for ever more constrained
architectures.

With the above motivation in mind, we proceed to the discussion of
coupling superconducting flux qubits.  Coherent oscillations of the
state of a superconducting flux qubit were first demonstrated at
Delft in 2003 \cite{Chio03}. A number of other institutions are also
developing flux qubit technology, including Berkeley \cite{Hime06},
NEC \cite{Nisk06,Nisk06b}, NTT \cite{Kaku07} and IPHT \cite{Graj05}.
A flux qubit is essentially a superconducting ring interrupted by
typically three Josephson junctions with clockwise and anticlockwise
persistent currents forming the basis of an effective two level
quantum system. For an up-to-date review of superconducting qubit
theory in general, including flux qubits, see \cite{Gell06}.

\begin{figure}
\begin{center}
\resizebox{55mm}{!}{\includegraphics{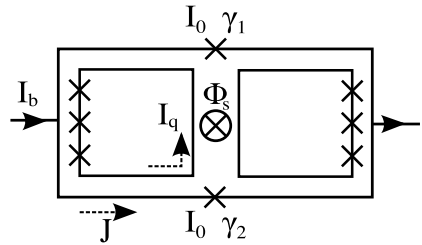}}
\end{center}
\caption{Coupling scheme as proposed in \cite{Plou04}, including
circuit symbols and orientations.} \label{simpCoup}
\end{figure}

The coupling scheme as proposed in \cite{Plou04} is shown, somewhat
simplified, in Fig.~\ref{simpCoup}.  The strength of the direct
inductive coupling between the qubits is given by $K_{0} =
2M_{qq}I_{q}^{2}$, whereas that mediated by the SQUID takes the form
\begin{equation}
K_{s} = 2M_{qs}^{2}I_{q}^{2}\left( \frac{\partial J}{\partial
\Phi_{s}}\right)_{I_{b}}.
\end{equation}
In the slowly varying and high resistance limit of the junctions, we
can write
\begin{eqnarray}
\label{currenteqns}
I_{b} & = & I_{0}\sin{\gamma_{1}} + I_{0}\sin{\gamma_{2}}, \\
2J & = & I_{0}\sin{\gamma_{2}} - I_{0}\sin{\gamma_{1}}
\end{eqnarray}
which can also be written as
\begin{eqnarray}
\label{currenteqns2}
I_{b} & = & 2I_{0}\sin{\bar\gamma}\cos{\Delta\gamma}, \\
\label{currenteqns22} J & = &
I_{0}\sin{\Delta\gamma}\cos{\bar\gamma}
\end{eqnarray}
where $\Delta\gamma = \frac{\gamma_{2} - \gamma_{1}}{2}$ and
$\bar{\gamma} = \frac{\gamma_{1} + \gamma_{2}}{2}$. These equations
are constrained by
\begin{equation}
\label{constraint1} \Delta\gamma = \frac{\pi}{\Phi_{0}}(\Phi_{s} -
LdJ)
\end{equation}
where $L$ is the inductance of the coupler and $\Phi_{s}$ is the
applied flux, nominally set to $\Phi_{s}=0.45\Phi_{0}$ to maximize
the response of $J$ to variations in $\Phi_{s}$. Taking the partial
derivative of Eqs.~(\ref{currenteqns2}), (\ref{currenteqns22}), and
(\ref{constraint1}) with respect to $\Phi_{s}$ we obtain
\begin{equation}
\frac{\partial I_{b}}{\partial\Phi_{s}} = 0 =
2I_{0}\frac{\partial}{\partial\Phi_{s}}\left(\cos{\Delta\gamma}\sin{\bar{\gamma}}\right),
\end{equation}
\begin{equation}
\frac{\partial J}{\partial\Phi_{s}} = I_{0}\left[ -\frac{\partial
\bar{\gamma}}{\partial\Phi_{s}}\sin{\bar{\gamma}}\sin{\Delta\gamma}
+ \frac{\partial
\Delta\gamma}{\partial\Phi_{s}}\cos{\bar{\gamma}}\cos{\Delta\gamma}\right],
\end{equation}
\begin{equation}
\frac{\partial\Delta\gamma}{\partial\Phi_{s}} =
\frac{\pi}{\Phi_{0}}(1 - L\frac{\partial J}{\partial\Phi_{s}}).
\end{equation}
Using these equations, we can derive the expression
\begin{equation}
\label{transfer:function} \left(\frac{\partial
J}{\partial\Phi_{s}}\right)_{I_{b}} = \frac{1}{2L_{j}}\frac{ 1 -
\tan^2{\bar{\gamma}}\tan^2{\Delta\gamma}}{ 1 +
\frac{L}{2L_{j}}\left( 1 -
\tan^2{\bar{\gamma}}\tan^2{\Delta\gamma}\right)}
\end{equation}
where $L_{j} = \Phi_{0} /\left(2\pi I_{0}
\cos{\Delta\gamma}\cos{\bar{\gamma}}\right)$ is the Josephson
inductance. This expression characterizes the tunable nature of the
coupling scheme.

\begin{figure}
\begin{center}
\resizebox{70mm}{!}{\includegraphics{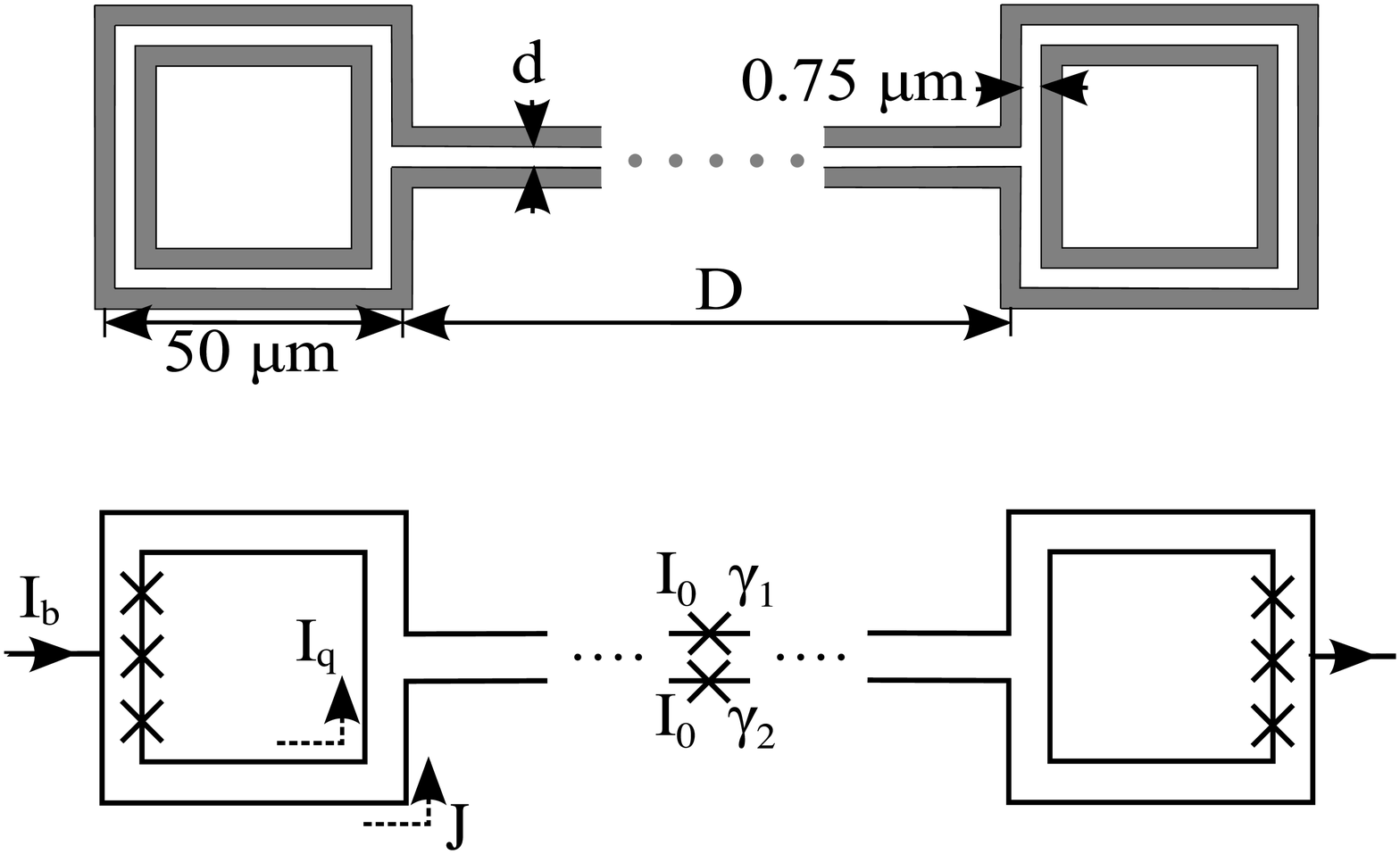}}
\end{center}
\caption{Extending coupling scheme, including circuit symbols,
orientations and dimensions.  All wires are 0.75 $\mu$m wide and 0.1
$\mu$m thick} \label{dogbone}
\end{figure}

We propose modifying the coupler as shown in Fig.~\ref{dogbone}. All
dimensions are typical of the Berkeley group \cite{Hime06}.  This
design results in decreased mutual inductance between the qubits,
increased mutual inductance between the qubits and the coupler,
increased self inductance of the coupler and significant capacitance
structurally incorporated into the coupler.  All of these effects
will be investigated, and the resultant impact on the coupling
strength.

\begin{figure}
\begin{center}
\resizebox{70mm}{!}{\includegraphics{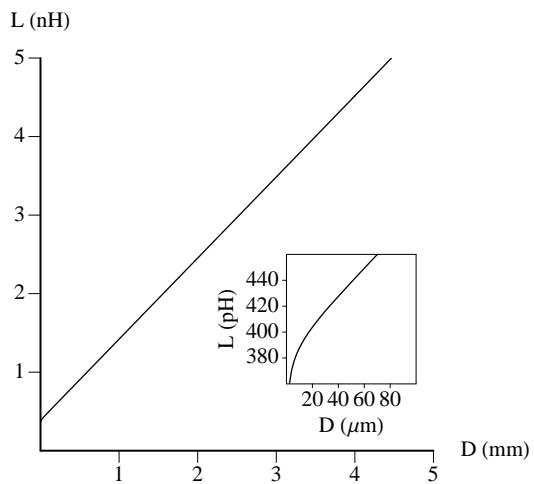}}
\end{center}
\caption{The self-inductance of the coupler as a function of the
length of the coupler.} \label{indVlen}
\end{figure}

Before discussing coupling strengths, we need to determine the
various inductances of the new system.  Fig.~\ref{indVlen} shows the
self inductance of the coupler versus coupler length $D$ for coupler
width $d=1.5$ $\mu$m generated using FastHenry assuming
superconducting aluminum wires with a penetration depth of 49 nm
\cite{Fabe55}. The inset shows the short length behavior. For all
lengths of interest, the coupler self inductance is approximately
given by ($356 + 0.863 D/\mu$m) pH. A small value of $d$ is
desirable to minimize $L$ at a given length and, as we shall see,
maximize the coupling strength. However, this also introduces a
large capacitance, with consequences to be discussed at the end of
this section.  The mutual inductance of each qubit with the coupler
was found to be 75 pH.

\begin{figure}
\begin{center}
\resizebox{70mm}{!}{\includegraphics{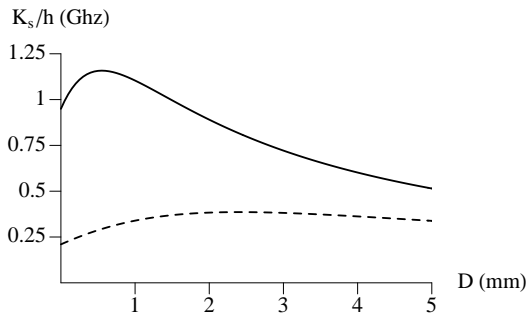}}
\end{center}
\caption{The coupling strength at zero bias current
$(\Phi_{s}=0.45\Phi_{0})$ without mutual qubit interaction versus
the length of the coupler for $I_{0} = 0.48$ $\mu$A (solid) and
$I_{0} = 0.16$ $\mu$A (dashed).} \label{strVlen}
\end{figure}

\begin{figure}
\begin{center}
\resizebox{70mm}{!}{\includegraphics{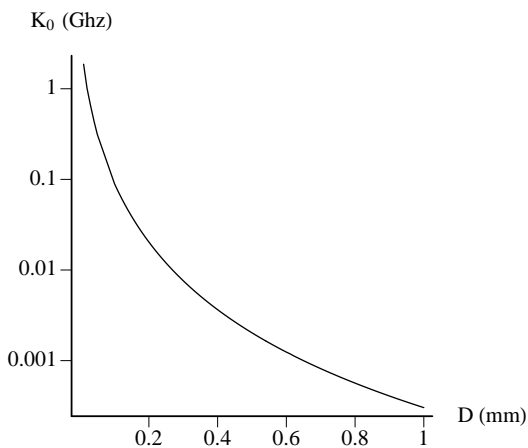}}
\end{center}
\caption{The strength of the direct qubit-qubit interaction due to
mutual inductive coupling as a function of the length of the
coupler, and thus qubit separation.} \label{indVlendirect}
\end{figure}

We are now in a position to calculate the coupler mediated coupling
strength $K_{s}$ for zero bias current, done numerically for two
different critical currents $I_0=0.48$ and $0.16$ $\mu$A and shown
in Fig.~\ref{strVlen}. Note the existence of optimum lengths, 700
and 2500 $\mu$m respectively, a consequence of the cosine terms in
$L_{j}$.  The coupling strength due to the mutual inductance of the
qubits is shown in Fig.~\ref{indVlendirect}. Note that to neglect
the direct coupling it is necessary for the coupler to be greater
than approximately 650 $\mu$m long, corresponding to a coupling
strength approximately 3 orders of magnitude less than that mediated
by a short coupler. The consequences of this for architecture design
are discussed in Section~\ref{archit}.

\begin{figure}
\begin{center}
\resizebox{70mm}{!}{\includegraphics{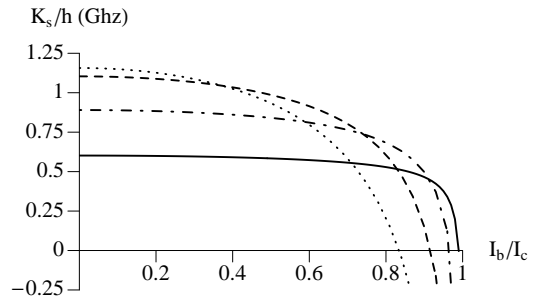}}
\end{center}
\caption{The coupling strength as a function of the bias current
with $\Phi_{s}=0.45\Phi_{0}$ and $I_{c}=I_{c}(\Phi_{s})$, using
Josephson Junction critical currents of $I_{0} = 0.48$ $\mu$A. $D =
500$ $\mu$m (dotted), $D = 1000$ $\mu$m (dashed), $D = 2000$ $\mu$m
(dash-dotted), and $D = 4000$ $\mu$m (solid).} \label{strVcurr1}
\end{figure}

\begin{figure}
\begin{center}
\resizebox{70mm}{!}{\includegraphics{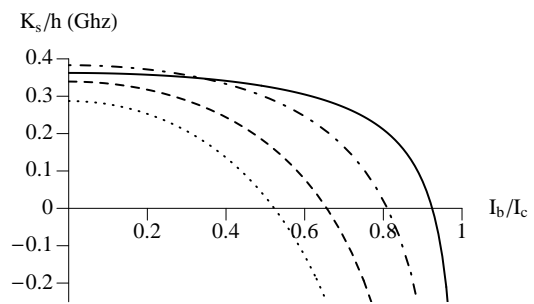}}
\end{center}
\caption{The coupling strength as a function of the bias current
with $\Phi_{s}=0.45\Phi_{0}$ and $I_{c}=I_{c}(\Phi_{s})$, using
Josephson Junction critical currents of $I_{0} = 0.16$ $\mu$A. $D =
500$ $\mu$m (dotted), $D = 1000$ $\mu$m (dashed), $D = 2000$ $\mu$m
(dash-dotted), and $D = 4000$ $\mu$m (solid).} \label{strVcurr2}
\end{figure}

The coupling strength can be reduced to zero by sufficiently
increasing the bias current.  It is desirable to ensure that the
necessary increase is as small as possible, as the presence of a
bias current, particularly one close to the critical current, is a
significant source of decoherence \cite{Plou04,Nisk06,vdWal03}.
Fig.~\ref{strVcurr1} shows the coupling strength versus bias current
for a selection of four coupler lengths.  This figure uses the
critical current $I_{0} = 0.48$ $\mu$A from \cite{Plou04}.  Clearly,
particularly for long lengths, the bias current required to achieve
zero coupling strength is too close to the critical current.  This
problem can be circumvented by reducing the critical current of the
junctions to $I_{0} = 0.16$ $\mu$A resulting in
Fig.~\ref{strVcurr2}. Reducing the critical current reduces the zero
bias coupling strength lowering the ratio $I_{b}/I_{c}$ required to
achieve zero coupling strength.

In principle, Fig.~\ref{strVcurr2} is promising, both in terms of
coupling strength and coupling length.  However, the effect of the
capacitance of the coupler must be determined.  As a starting point,
consider a single flux qubit initially prepared in a clockwise
current state.  Left alone, this qubit will oscillate between
clockwise and anticlockwise current states at its tunneling
frequency, which is typically of order a few GHz in current devices.
As seen by the coupler, by virtue of their mutual inductance, such a
qubit plays the role of an alternating current source.  Considering
the coupler in isolation now, we wish to check that an alternating
current source at one end of the coupler with amplitude $A$
generates an alternating current of amplitude as close to $A$ as
possible at the other end. Using a lumped circuit model and
discretizing the capacitive section of the coupler, deviation from
perfect transmission of order 1\% was found.  This is low enough to
give us confidence that the fundamental concept of the extended
coupler is sound, but high enough that achieving high fidelity gates
will require a closer examination of the physics of the system
\cite{Hutt06}. We have also begun simulations of the complete system
including silicon substrate using the commercial package HFSS.  The
results of these simulations and capacitive and radiative effects in
general will be discussed in detail in a separate publication.

\section{Architectures}
\label{archit}

\begin{figure}
\begin{center}
\resizebox{70mm}{!}{\includegraphics{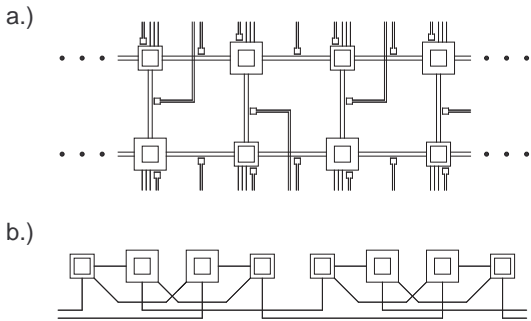}}
\end{center}
\caption{a) Simple scalable bilinear flux qubit architecture
including flux lines for each coupler and qubit. As drawn, the
architecture requires three layers of metal corresponding to the
three couplers around each qubit. b) Topologically identical array
used as the basis of Fig.~\ref{complete}.} \label{two_base_archs}
\end{figure}

In quantum computing literature, the word ``scalable'' is,
regrettably, frequently used rather loosely, and sometimes
inaccurately. Ideally, as a minimum, it should only be claimed that
an architecture is scalable if in principle an arbitrarily large
number of qubits can be implemented, the number of quantum gates and
measurements that can be executed simultaneously grows linearly with
the number of qubits, and the physics of any one quantum gate or
measurement does not depend on the total number of qubits. A simple
example of such an architecture making use of the coupler described
in this paper is shown, not drawn to scale, in
Fig.~\ref{two_base_archs}a. This architecture requires three
couplers around each qubit, well within current layering technology.
Note that one qubit of each pair is more weakly coupled following
\cite{Hime06}.  This enables readout of both qubits simultaneously
using one coupler via the resonant readout scheme of \cite{Serb07}.
Given the large potential length of each coupler, crosstalk can be
neglected, and with only a small number of control lines leading to
external circuitry per pair of qubits, given a sufficiently large
fridge, many qubits can be accommodated. The details of how many
qubits and how to include the necessary control lines and classical
control machinery shall be left for a future publication. Two lines
of qubits have been incorporated in the design to permit simpler
error correction resulting in a threshold two-qubit gate error rate
for arbitrarily large fault-tolerant quantum computation of
$1.96\times 10^{-6}$ as described in \cite{Step07a}. Note that to
use this architecture in practice this implies the need for
two-qubit gates operating with an error rate of $10^{-7}$ or less,
far below what is achievable in most solid-state systems given the
current ratios of decoherence times to gate times.

\begin{figure}
\begin{center}
\resizebox{70mm}{!}{\includegraphics{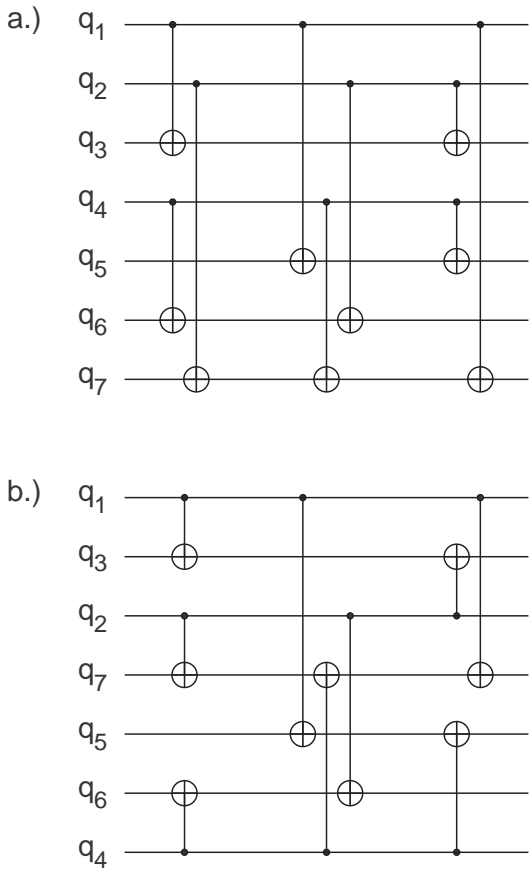}}
\end{center}
\caption{To reduce the need for long range interactions, the circuit
a.) taken from \cite{DiVi06}, which encodes and decodes a logical
zero, was modified as shown in b.) by swapping two pairs of qubits.}
\label{zero}
\end{figure}

\begin{figure*}
\begin{center}
\resizebox{115mm}{!}{\includegraphics{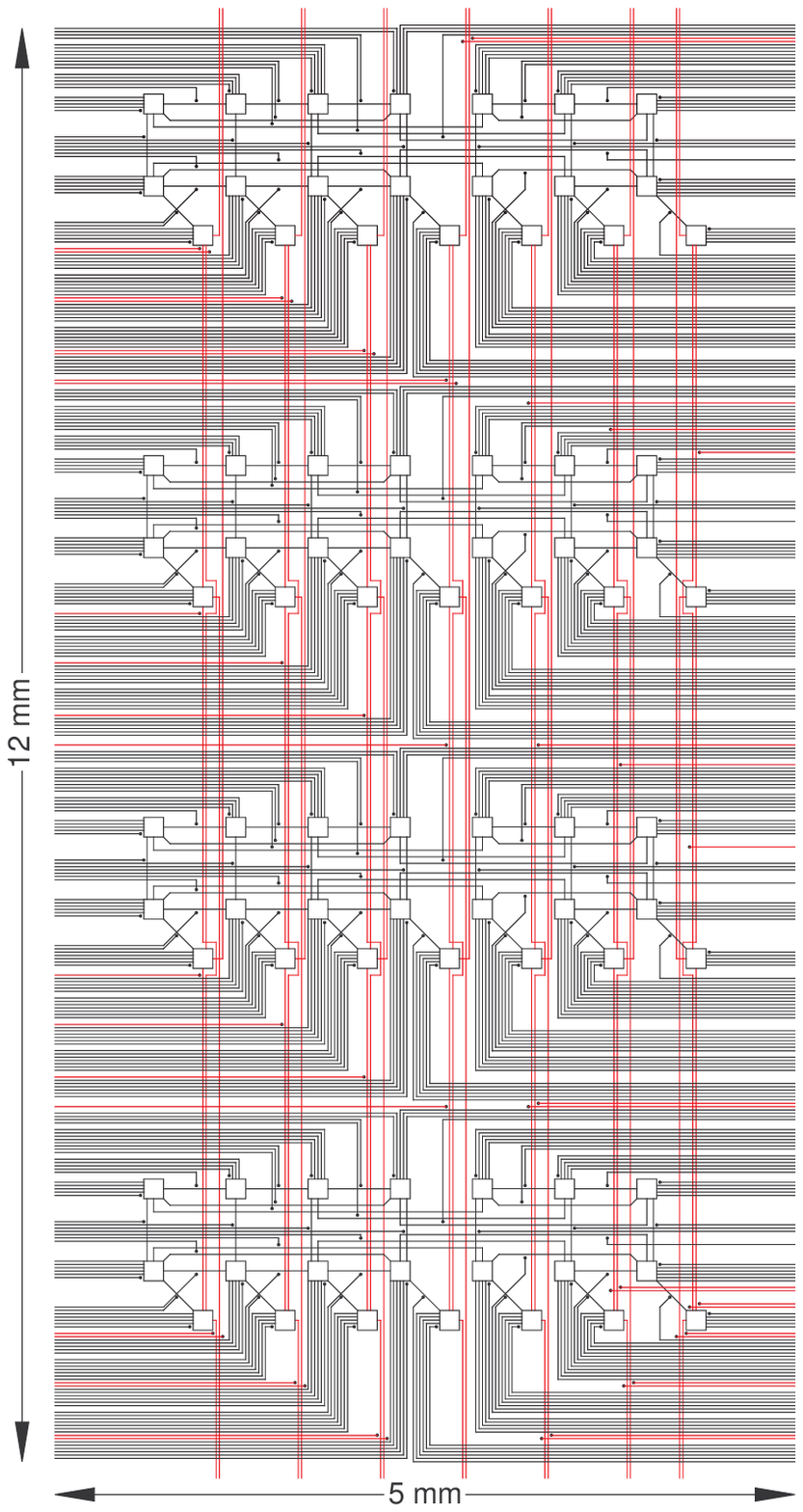}}
\end{center}
\caption{A complete architecture for a flux qubit quantum computer.
Squares represent flux qubits, lines connecting squares represent
SQUID couplers, lines running away from each square represent squid
bias current lines and flux bias lines. The size of the flux qubits
has been exaggerated for clarity. Lines colored red represent the
SQUIDs and flux bias lines for the second and higher levels of error
correction and logical gates.} \label{complete}
\end{figure*}

Of course, the architecture of Fig.~\ref{two_base_archs}a does not
take full advantage of the potential length of the coupler.  As
discussed earlier, to ensure relatively low crosstalk, from
Fig.~\ref{indVlen}, qubits need to be spaced approximately 650
$\mu$m apart.  This still gives us enough space to firstly stretch
the architecture of Fig.~\ref{two_base_archs}a into a single line as
shown in Fig.~\ref{two_base_archs}b, then duplicate this line 7
times to permit an additional layer of the 7-qubit Steane code to be
used.  Referring to Fig.~\ref{complete}, these duplicated qubits
correspond to the bottom seven qubits in each group of 21.  The
middle row of qubits in each group of 21 correspond to ancilla
qubits used during error correction according to the scheme
described in \cite{DiVi06}, with slight modifications to reduce the
range of the necessary interactions as shown in Fig.~\ref{zero}. The
top row of qubits in each group of 21 correspond to ancilla qubits
only used during the implementation of the fault-tolerant T gate, or
$\pi/8$ gate, as it is also known, which is required to ensure the
computer can perform a universal set of fault-tolerant gates.  A
complete description of the circuitry of this gate can be found in
\cite{Step07a}.  Note that, even with all the additional control
lines, the architecture requires just one wire per approximately 40
$\mu$m per side.

The network of qubits and couplers shown in Fig.~\ref{complete}
enables the most efficient known nonlocal error correction scheme
and fault-tolerant gates to be implemented at the lowest level.  All
higher levels make use of the circuitry devised for the bilinear
architecture.  When the threshold two-qubit error rate of this more
complicated architecture was calculated, using the mathematical
tools described in \cite{Step07a}, the disappointingly low result of
$6.25\times 10^{-6}$ was obtained
--- just over a factor of three better than the bilinear array.  In
short, the dominant nearest neighbor behavior of the large-scale
architecture is not overcome by a single layer of nonlocal error
correction and gates.

\section{Conclusion}
\label{conc}

We have described in detail a nonlocal method of coupling pairs of
flux qubits and shown that it is suited to the construction of
complex, scalable quantum computer architectures as shown in
Fig.~\ref{complete}. By virtue of the fact that flux qubits do not
require large quantities of classical control circuitry on chip,
there are no obvious lithographic or heat dissipation barriers to
the construction of such an architecture.  The primary concern, as
with all superconducting quantum technology, is decoherence.  In the
near future we wish to look at other superconducting qubits and
coupling schemes with the aim of removing all known sources of
decoherence from the design.  For example, in \cite{Nisk06} a method
of coupling flux qubits tunably is described that does not resort to
a nonzero SQUID bias current.  Furthermore, the need for large qubit
separations to minimize crosstalk could in principle be alleviated
by enclosing each qubit in a micrometer scale Meissner cage.
Devising a more practical method of achieving higher qubit densities
would greatly increase the utility of the proposed coupling scheme.
Finally, with or without higher qubit densities, additional
architecture design work is required to try to raise the threshold
further, possibly by attempting to incorporate two layers of
nonlocal error correction into the design.

\section{Acknowledgements}

AGF and FKW would like to thank S. Oh for helpful discussions. This
work was supported by an NSERC Discovery Grant and the European
Union through EuroSQIP.

\bibliography{../References} % Produces the bibliography via BibTeX.

\end{document}